\documentclass[journal]{IEEEtran}
\usepackage[left=1.50cm, right=1.50cm, top=1.5cm, bottom = 1.50cm]{geometry}
\IEEEoverridecommandlockouts
\usepackage{cite}
\usepackage{subfigure}
\usepackage{bbm}
\usepackage{amsmath,amssymb,amsfonts}
\usepackage[ruled,vlined,linesnumbered]{algorithm2e}
\usepackage{graphicx}
\usepackage{algpseudocode}
\usepackage{epstopdf}
\usepackage{caption}
\usepackage{textcomp}
\usepackage{soul,color}
\usepackage{epstopdf}
\usepackage{ragged2e}
\usepackage[utf8]{inputenc}
\usepackage[T1]{fontenc}
\usepackage{mwe}
\usepackage{soul}
\usepackage[utf8]{inputenc}
\usepackage{url}
\usepackage[utf8]{inputenc}
\usepackage[english]{babel}
\usepackage{amsthm}
\usepackage[symbol]{footmisc}
\theoremstyle{definition}

\usepackage{lscape}
\usepackage{rotating}
\usepackage{tabularx}
\theoremstyle{remark}

\soulregister\cite7
\soulregister\ref7
\soulregister\pageref7
\def\BibTeX{{\rm B\kern-.05em{\sc i\kern-.025em b}\kern-.08em
    T\kern-.1667em\lower.7ex\hbox{E}\kern-.125emX}}

\DeclareMathOperator*{\E}{\mathbb{E}}

\usepackage{xcolor,colortbl}

\definecolor{Gray}{gray}{0.8}

\newcolumntype{a}{>{\columncolor{Gray}}c}
\newcolumntype{b}{>{\columncolor{white}}c}
\makeatletter

\def\therule{\makebox[\algorithmicindent][l]{\hspace*{.5em}\vrule height .75\baselineskip depth .25\baselineskip}}%

\newtoks\therules
\therules={}
\def\appendto#1#2{\expandafter#1\expandafter{\the#1#2}}
\def\gobblefirst#1{
	#1\expandafter\expandafter\expandafter{\expandafter\@gobble\the#1}}%
\def\LState{\State\unskip\the\therules}
\def\pushindent{\appendto\therules\therule}%
\def\popindent{\gobblefirst\therules}%
\def\printindent{\unskip\the\therules}%
\def\printandpush{\printindent\pushindent}%
\def\popandprint{\popindent\printindent}%

\algdef{SE}[WHILE]{While}{EndWhile}[1]
{\printandpush\algorithmicwhile\ #1\ \algorithmicdo}
{\popandprint\algorithmicend\ \algorithmicwhile}%
\algdef{SE}[FOR]{For}{EndFor}[1]
{\printandpush\algorithmicfor\ #1\ \algorithmicdo}
{\popandprint\algorithmicend\ \algorithmicfor}%
\algdef{S}[FOR]{ForAll}[1]
{\printindent\algorithmicforall\ #1\ \algorithmicdo}%
\algdef{SE}[LOOP]{Loop}{EndLoop}
{\printandpush\algorithmicloop}
{\popandprint\algorithmicend\ \algorithmicloop}%
\algdef{SE}[REPEAT]{Repeat}{Until}
{\printandpush\algorithmicrepeat}[1]
{\popandprint\algorithmicuntil\ #1}%
\algdef{SE}[IF]{If}{EndIf}[1]
{\printandpush\algorithmicif\ #1\ \algorithmicthen}
{\popandprint\algorithmicend\ \algorithmicif}%
\algdef{C}[IF]{IF}{ElsIf}[1]
{\popandprint\pushindent\algorithmicelse\ \algorithmicif\ #1\ \algorithmicthen}%
\algdef{Ce}[ELSE]{IF}{Else}{EndIf}
{\popandprint\pushindent\algorithmicelse}%
\algdef{SE}[PROCEDURE]{Procedure}{EndProcedure}[2]
{\printandpush\algorithmicprocedure\ \textproc{#1}\ifthenelse{\equal{#2}{}}{}{(#2)}}%
{\popandprint\algorithmicend\ \algorithmicprocedure}%
\algdef{SE}[FUNCTION]{Function}{EndFunction}[2]
{\printandpush\algorithmicfunction\ \textproc{#1}\ifthenelse{\equal{#2}{}}{}{(#2)}}%
{\popandprint\algorithmicend\ \algorithmicfunction}%
\makeatother

\begin{document}
		\title{\huge Optimizing Age of Information Through Aerial Reconfigurable Intelligent Surfaces: A Deep Reinforcement Learning Approach \\	
	}	

	\author{
	\IEEEauthorblockN{Moataz Samir\textsuperscript{1}, Mohamed Elhattab\textsuperscript{1}, Chadi Assi\textsuperscript{1}, Sanaa Sharafeddine\textsuperscript{2}, and Ali Ghrayeb\textsuperscript{3}}\\
		\IEEEauthorblockA{\textit{\textsuperscript{1}Concordia University,  \textsuperscript{2}Lebanese American University, \textsuperscript{3}Texas A\&M University at Qatar}\\
	}}
	\maketitle
	
\begin{abstract}
We investigate the benefits of integrating unmanned aerial vehicles (UAVs) with reconfigurable intelligent surface (RIS) elements to passively relay information sampled by Internet of Things devices (IoTDs) to the base station (BS). In order to maintain the freshness of relayed information, an optimization problem with the objective of minimizing the expected sum Age-of-Information (AoI) is formulated to optimize the altitude of the UAV, the communication schedule, and phases-shift of RIS elements. In the absence of prior knowledge of the activation pattern of the IoTDs, proximal policy optimization algorithm is developed to solve this mixed-integer non-convex optimization problem. Numerical results show that our proposed algorithm outperforms all others in terms of AoI.

\end{abstract}

\section{Introduction}
Numerous emerging smart city applications rely on freshness of sensory data (i.e., status-updates) which is being monitored and generated by a plethora of Internet of Things devices (IoTDs). For instance, smart environmental monitoring, industrial control systems and intelligent transportation systems all require reliability and timeliness in delivering status-update information. Outdated updates may be inconsistent with the current status of the physical process being monitored and controlled, which may lead to erroneous decisions. Despite their great benefits, IoTDs have limited capabilities and cannot communicate over longer distances in a reliable manner. As a result, providing a timely and reliable communication service for IoT devices is a challenging task, which may hinder their expected benefits. Latency in delivering sampled information consists of three main delays: the delay until data is being generated, the delay until the generated update is scheduled for transmission, and the intermediate processing delay between the source and the destination. In order to fully characterize the freshness of status-update, the concept of \textit{Age-of-Information} (AoI) has been introduced as a main performance metric for these types of applications. AoI can be defined as the elapsed time since the generated/sampled of the most a recently received status-update. Undoubtedly, emerging IoT services will strongly benefit from enhancing wireless connectivity, which is considered as an enabler for the evolution of future networks and their services. While many key enabling technologies are considered to unleash the potential of future networks, a revolutionary one (which exploits the radio environment has a new degree of freedom) has recently emerged and is under intense investigation. 

\par Reconfigurable Intelligent Surfaces (RISs) leverage the tuning capabilities of their reflective elements to enhance the propagation environment by improving the desired signal at the receiver and mitigating interference. They are energy efficient and expected to greatly enhance the spectral efficiency of wireless networks, particularly when combined with other promising technologies. UAVs are among those technologies that have shown great promise in assisting networks by improving connectivity and coverage. Unlike UAVs, as conventional mobile relaying elements, integrating reconfigurable intelligent surfaces with the UAVs allows several benefits. First, the data transmission from the IoTDs to the BS through RIS-empowered UAV will experience less intermediate delays to relay the information compared to UAVs acting as mobile active relays. This is because, in a decode and forward half-duplex relaying mode, the transmission is executed over two time slots. On the other hand, an IRS-integrated UAV, which is denoted as aerial RIS (ARIS), requires only one time slot since the RIS operates in a full-duplex (FD) relaying mode. This may enhance the freshness of information and lead to reducing the AoI. Second, the power consumption due to processing the relayed information at the UAVs can be avoided, which increases the flight endurance of the UAV. In fact, RIS is composed of a large number of passive low-cost elements, each of which is capable of independently tuning the phase-shift of the incident radio waves \cite{Elhattab}. For instance, by appropriately configuring the phase-shift with the aid of the RIS controller, the reflected signals can be constructively added, and therefore, enhancing the reliability of IoT networks, accordingly, the AoI is minimized. 

\par Recently, some efforts have  been  directed towards the integration between the UAVs and the RIS with optimizing and designing the phase-shift of the RIS elements in order to intensify different utilities such as the network spectral efficiency \cite{RIS_SE}, network energy efficiency \cite{RIS_EE}, max-min fairness \cite{RIS_Fairness}, the coverage probability \cite{RIS_Coverage}, or the power consumption \cite{RIS_Power}. However, these solutions may not be necessarily optimal from the perspective of preserving freshness of information. To the best of our knowledge, none of the previous works reported in the literature has addressed the optimization of RIS configuration while considering the freshness of information, which thus motivates this work.

\par We study a wireless network where IoTDs with limited transmission capabilities sample a stochastic process and the sampled data needs to be processed by a BS. A single UAV equipped with RIS is deployed to act as a passive relay node to forward the sampled data to the BS while considering the different activation patterns of IoTDs. The sampled data is successfully delivered to the BS if and only if the signal-to-noise ratio (SNR) exceeds a predefined threshold, upon which the AoI decreases. This framework is formulated as an optimization problem with the objective of minimizing the expected sum AoI (ESA) while considering the SNR constraints, UAV altitude constraint, and the IoTDs scheduling constraints. Then the optimization problem is shown to be difficult to solve while considering a realistic challenging scenario where the activation patterns of IoTDs are unknown. Therefore, we opt to apply a deep reinforcement learning (DRL) framework based on proximal policy optimization (PPO) to learn randomness of the IoTDs' activation patterns and control the altitude of the UAV, the phase-shift of RIS elements along with communication scheduling to minimize the ESA. 

\section{System Model and Problem Formulation} \label{SystemMOverviewodelCommunication}

\begin{figure}[t!]
	\includegraphics[width=0.4\textwidth]{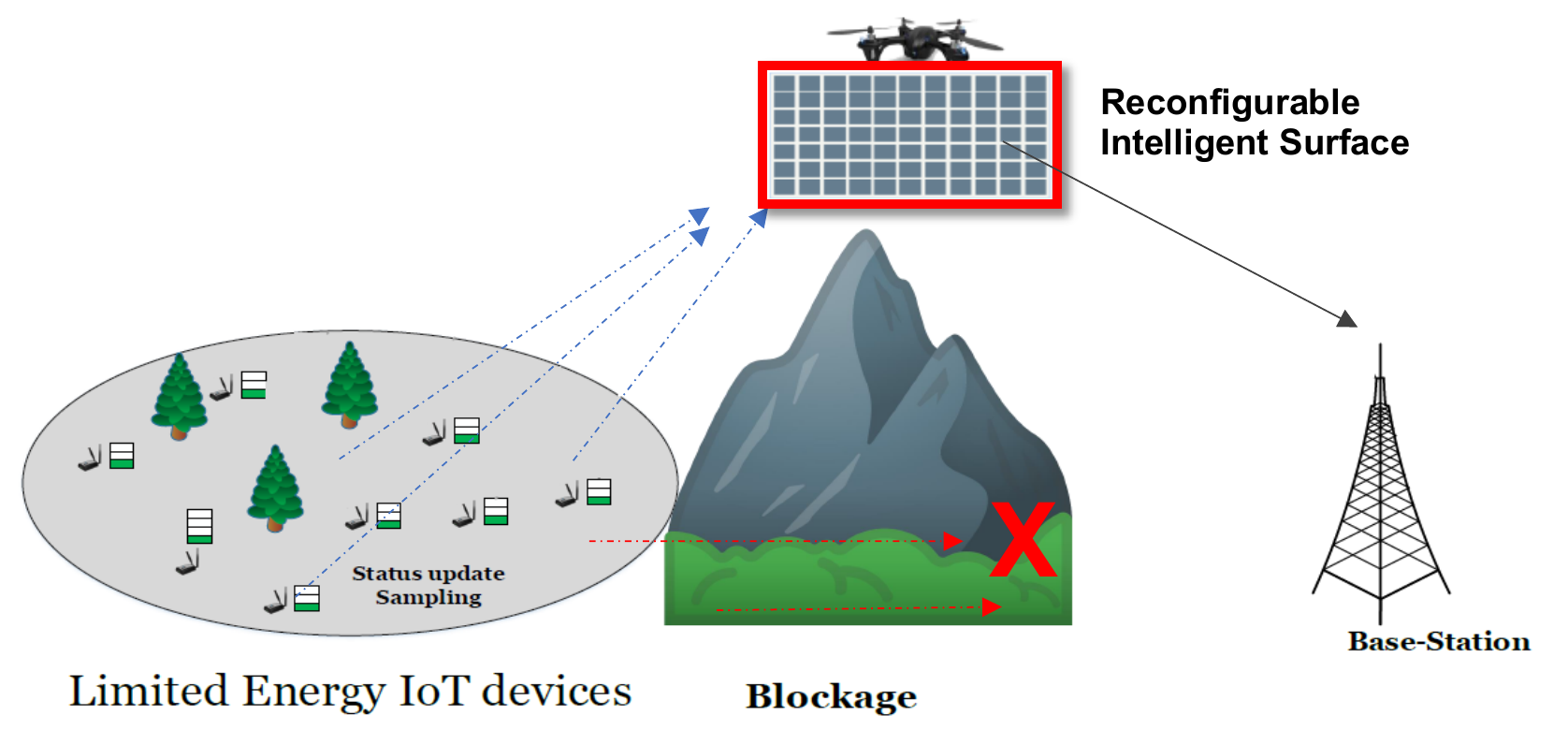}
	\centering
	\caption{ARIS-assisted IoT wireless networks}
	\centering
	\label{Fig:Maievevn}
\end{figure}

As illustrated in Fig.\ref{Fig:Maievevn}, we consider an IoT wireless network where a set $\cal{M}$ of $M$ IoTDs with limited capabilities are deployed to provide time-stamped, status-update information. Due to IoTDs' capabilities constraints and environmental obstacles, the existence of a strong direct line-of-sight (LoS) communication link is difficult to obtain. Therefore, a single UAV equipped with RIS consisting of $F$ reflecting elements is deployed to passively relay the status-update information to the BS. We consider the system over multiple time frames. Each of these frames is further divided into equal segments, that is, $N$ time-slot of length $\delta_{t}$, which is normalized to unity.


\par The planar coordinates of the deployed UAV are assumed to be placed at $(x_U, y_U)$. In Cartesian coordinates, the locations of IoTDs are assumed to be known and located at $(x_i,y_i, 0), \forall i \in M$ at ground level. Depending on the services/applications, the activation patterns for IoTDs are different\footnote{As mentioned in the 3rd generation partnership project (3GPP).}. In addition, we assume that the BS is located at $(x_s, y_s, H_S)$, where $H_S$ denotes the height of the BS. At any given time-slot $n$, the deployed UAV can adapt its altitude $H_U[n]$ such that $H_U[n] \in (H_{min}, H_{max})$, where $H_{min}$ and $H_{max}$ are the minimum and the maximum altitude range specified by aviation authorities, respectively. Consequently, the altitude control should meet the following constraints:
\begin{equation}
	 H_{min} \le H_U[n] \le H_{max}, \forall n,	
	\label{eq11}
\end{equation}
\begin{equation}
	{\Big|H_{U}[n+1]- H_{U}[n]\Big|} \leq D_{\max}, n = 1, ...,N - 1,
	\label{eq22}
\end{equation}
\begin{equation}
	 H_{U}[1] = H_{S},
	\label{eq33}
\end{equation}
where $D_{\max}= V_{\max}\delta_{t}$ is the maximum vertical distance by the UAV in one time-slot based on its maximum speed $V_{\max}$ and $H_{S}$ denotes initial vertical location.
\par The BS continuously controls the altitude of the UAV as well as the phase-shift of the reflecting elements in order to serve the IoTDs and maintain their required quality of service (QoS). Let $\boldsymbol{\Phi}[n] =$ diag$\{e^{j\phi_1[n]}, e^{j\phi_2[n]}, . . .e^{j \phi_f[n]}, ....  ,e^{j\phi_F[n]}\} \in \mathbb{C}^{F\times F}$ be the RIS's diagonal phase-shift matrix in the $n$th time-slot, where $\phi_f[n] \in [0, 2\pi), ~ \forall f \in F$ is the phase-shift for the $f$th reflecting element. With time division multiple access, the UAV schedules at most one IoTD to transmit its status-update. Therefore, the transmission scheduling should meet the constraint below:
\begin{equation}
\sum_{i=1}^ {M} \alpha_{i}[n] \leq 1, ~~~\forall n, 
\label{eq:226263dcd22}
\end{equation}
where $\alpha_{i}[n]$ is a binary variable, which indicates that IoTD $i$ is scheduled in time-slot $n$, and $0$ otherwise. 
\par Before we proceed, we define the distance model and the adopted channel gain model. We denote $d_{i \rightarrow U}[n]$ and $d_{U \rightarrow S}[n]$ as the distance between the IoTDs and the UAV in the $n$th time-slot and between the UAV and the BS, respectively, which are given as follows.
\begin{equation}
	d_{i \rightarrow U}[n]={
		\sqrt{(x_{i}-x_U)^{2}+(y_{i}-y_U)^{2}+(H_U[n])^{2}}}, 
	\label{eq4}
\end{equation}
and 
\begin{equation}\small
	d_{U \rightarrow S}[n]={
		\sqrt{(x_S-x_U)^{2}+(y_S-y_U)^{2}+(H_S-H_U[n])^{2}}}.
	\label{eq34}
\end{equation}
\par Meanwhile, the channel gain between the IoTDs and the UAV, and between the UAV and the BS are denoted as $h_{i \rightarrow U}[n]$ and $h_{U \rightarrow S}[n]$ and can be expressed as follows,
\begin{equation}
h_{i \rightarrow U}[n]= \widehat{h}_{i \rightarrow U}[n] \Delta_{i \rightarrow U}[n], 
\label{eq:3399874443}
\end{equation}
and
\begin{equation}
h_{U \rightarrow S}[n]= \widehat{h}_{U \rightarrow S}[n] \Delta_{U \rightarrow S}[n],
\label{eq:334443}
\end{equation}
where $\widehat{h}_{i \rightarrow U}[n]$, $\widehat{h}_{U \rightarrow S}[n]$, $\Delta_{i \rightarrow U}[n]$ and $\Delta_{U \rightarrow S}[n]$ represent the small-scale fading between the IoTDs and the UAV, the small-scale fading between the UAV and the BS, the path-loss coefficients between the IoTDs and the UAV, the path-loss coefficients between the UAV and the BS, respectively. In particular, the path-loss coefficients $\Delta_{i \rightarrow U}[n]$ and $\Delta_{U \rightarrow S}[n]$ can be written as $\Delta_{i \rightarrow U}[n]= \sqrt{\gamma_0 d^{-\eta}_{i \rightarrow U}[n]}$ and $\Delta_{U \rightarrow S}[n]= \sqrt{\gamma_0 d^{-\eta}_{U \rightarrow S}[n]}$,  where $\gamma_0$ is the path-loss average channel power gain at a reference distance $d_0 = 1m$, $\eta$ is the path-loss exponent. Similar to \cite{Rician_2, RIS_Coverage}, we consider a Rician fading with a dominant LoS. Thus, the small-scale fading for the communication link between the IoTD and the UAV $\widehat{h}_{i \rightarrow U}[n]$ can be given as follows
\begin{equation}
\widehat{h}_{i \rightarrow U}[n]= \sqrt{\dfrac{K_1}{K_1+1}} {\overline{\boldsymbol{h}}_{i \rightarrow U}[n]} ,
\label{eq:33nbnbn4443}
\end{equation}
where $K_1$ is the Rician factor. $\overline{\boldsymbol{h}}_{i \rightarrow U}[n] = \left[e^{j\psi_{i,1}}, e^{j\psi_{i,2}}, \dots e^{j\psi_{i,F}}\right]$ is a fixed component vector with elements
of unit power, and $\psi_{i, f} \in \left[0, 2\pi\right]$. Similarly, the small-scale fading for the communication link between the UAV and the BS $\widehat{h}_{U \rightarrow S}[n]$ can be given by
\begin{equation}
\widehat{\boldsymbol{h}}_{U \rightarrow S}[n]= \sqrt{\dfrac{K_2}{K_2+1}} {\overline{h}_{U \rightarrow S}[n]},
\label{eq:3344uuhuh43}
\end{equation}
 where $K_2$ is the Rician factor. $\overline{\boldsymbol{h}}_{U \rightarrow S}[n] = \left[e^{j\omega_{i, 1}}, e^{j\omega_{i, 2}}, \dots e^{j\omega_{i, F}}\right]$ is a fixed component vector with elements of unit power, and $\omega_{i,f} \in \left[0, 2\pi\right]$. All IoTDs are assumed to have the same transmit power denoted by $P$.

Based on the defined channel model in (\ref{eq4}) - (\ref{eq:3344uuhuh43}), the SNR at the BS in time-slot $n$ can be expressed as
\begin{equation}
	\Upsilon_{i}\Big(\boldsymbol{\Phi}[n] ,H_U[n]\Big)=\dfrac{P \Big|{\textbf{h}}^{H}_{i \rightarrow U}[n] \boldsymbol{\Phi}[n] {\textbf{h}}_{U \rightarrow S}[n]\Big|^2}{\sigma^2},
	\label{eq:334dvdvf443}
\end{equation}
where $\sigma^2$ is the thermal noise power. Note that, the overall channel gain between an IoTD and the BS, i.e., ${\textbf{h}}^{H}_{i \rightarrow U}[n] \boldsymbol{\Phi}[n] {\textbf{h}}_{U \rightarrow S}[n]$, can be written as
\begin{align}
&{\textbf{h}}^{H}_{i \rightarrow U}[n] \boldsymbol{\Phi}[n] {\textbf{h}}_{U \rightarrow S}[n] = \cr & \dfrac{\gamma_0^2 \sum^F_{f=1} |[\boldsymbol{h}_{U \rightarrow S}]_f| |[\boldsymbol{h}_{i \rightarrow U}]_f|~.~e^{j (\phi_f[n] - \psi_{i, f} - \omega_{i, f} )}}{{d^{-\eta/2}_{U \rightarrow S}[n]}{d^{-\eta/2}_{i \rightarrow U}[n]}},
	\label{eq:334devevvf443}
\end{align}
where $[\boldsymbol{h}]_f$ is the $f$th element of $\boldsymbol{h}$ and $\gamma_0$ is the path-loss average channel power gain at a reference equal to 1m.The maximum SNR at the BS can be achieved when the phase-shift is chosen as $\phi_f[n] = \psi_{i, f} + \omega_{i, f}$ \cite{Elhattab}.
\par  In order to achieve a successful transmission, $\Upsilon_{i}(\boldsymbol{\Phi}[n], H_U[n])$ should be strictly greater than or equal to $\Upsilon_{th}$, where $\Upsilon_{th}$ is the minimum threshold to ensure reliable decoding \cite{VPOORAGE}. 
%
%
%
A single packet queuing discipline is assumed to be employed by the IoTDs such that the older status-update packet is dropped and replaced with the newly arrived sample. A \textit{per time-slot sampling} policy is considered for sampling the information, where the scheduled activated IoTD samples the status-update information at the beginning of each time-slot to transmit its status-update information. Therefore, the deployed UAV has to control the scheduling, altitude and phases of RIS elements properly to relay the status-update information to the BS while considering the activation patterns of IoTDs. Clearly, AoI depends on the altitude of the UAV, the communication scheduling, phase-shift of the RIS elements and the activation pattern of the IoTDs. Thus, the evolution of $A_i[n]$ of IoTD $i$ can be written\footnote{For more tractable analysis, the initial values of AoI is neglected, that is,  $A_i[0]=0,~\forall i$.}
%
%
\begin{equation}\small
\begin{split}
	A_i[n+1]= 
	\begin{cases}
		 1,& \text{if } G_i[n]= 1 ,   \alpha_{i}[n]=1, \text{and} \\ & \Upsilon_{i} (\boldsymbol{\Phi}[n], H_U[n]) \geq \Upsilon_{th}, \\
		A_i[n]+1, & \text{otherwise},
	\end{cases}
	\end{split}
	\label{AOIequation}
\end{equation}
where $G_i[n]$ is a binary variable, which indicates that IoTD $i$ is active in time-slot $n$, and $0$ otherwise.
To obtain the AoI within the relay mission time, we use the  the ESA  $\dfrac{1}{NM} \mathbb{E} \Big[ \sum_{n=1}^N \sum_{i=1}^ M  A_i[n] | A_i[0]=0  \Big]$ as our metric to evaluate the freshness of sampled data. For the sake of tractability, the AoI can be expressed as the constraints below
\begin{equation}
	A_i[n+1]= 1+ A_i[n] -A_i[n] G_i[n] \alpha_{i}[n].
	\label{AOIgvgatsdsdion}
\end{equation}
\begin{equation}
	\Upsilon_{i}(\boldsymbol{\Phi}[n], H_U[n]) \geq G_i[n] \alpha_{i}[n] \Upsilon_{th}.
	\label{AOIgvgatisdson}
\end{equation}
%
With the quest of enhancing the performance of the IoTDs, a framework to optimize communication scheduling, phase-shift matrix of RIS, and the altitude of the UAV is investigated. This framework is formulated as an optimization problem with the objective of minimizing the ESA. For ease of notation, let us denote $\textbf{L} = \{H_U[n], \forall n\}$,  $\textbf{S} = \{\alpha_{i}[n], \forall i,n\}$ and  $\boldsymbol{\Theta} = \{\boldsymbol{\Phi}[n], \forall n\}$. Thus, our  problem can be written as:
\allowdisplaybreaks
\begingroup
\begin{subequations} \label{eq:subeqnsglobalEQ}
	\begin{align}
	\text{$\mathcal {OP}$:   }&	\min_{\substack{\textbf{L},\textbf{S}, \boldsymbol{\Theta} }} \dfrac{1}{NM}  \E \Big[ \sum_{n=1}^N \sum_{i=1}^ M   A_i[n] | A_i[0]=0  \Big] \\
	\text{s.t. } \,\,\,&  
	 (\ref{eq11})-(\ref{eq:226263dcd22}), (\ref{AOIgvgatsdsdion}),(\ref{AOIgvgatisdson}), \cr &
    \alpha_{i}[n]  \in \{0,1\}, ~~~\forall i,n,
    \label{eC1443}
	\\ &
	{\phi}_f[n] \in [0, 2\pi), ~~~\forall f, n,
	\label{eC22443}
	\end{align}
\end{subequations}
\endgroup

Owing to the randomness of the activation pattern, $G_i[n]$, of the IoTDs, $\mathcal {OP}$ is a stochastic optimization problem over the service time $N$. In fact, obtaining the activation patterns of IoTDs are crucial before dispatching the UAV to a target area. This is because the formulated problem aims to find the control policy that minimizes the AoI from the active IoTDs within the service time $N$. However, obtaining complete information on the activation pattern requires extensive measurement, which is not easy to obtained especially in remote areas. We also observe that $\mathcal {OP}$ is a mixed-integer  non-convex optimization problem which is hard to be solved. This is because $\mathcal {OP}$ contains both binary variables $\alpha_{i}[n]$ and continuous variables $\boldsymbol{\Phi}[n]$ and $H_U[n]$. In addition, it is a challenging task to solve a non-convex optimization problem in the absence of a complete information on the activation pattern. Therefore, our problem is reformulated as MDP and a model-free DRL based on PPO is exploited to find the effective control policy that minimizes the ESA. The proposed PPO algorithm does not rely on a prior knowledge of the activation patterns. 
%




~\vspace{-0.3in}
\section{Proposed Solution} \label{SysSolutiontAOI}

Finding the control policy that governs the UAV's altitude, the scheduling policy and the phase-shift matrix for the RIS is a non-trivial challenge. The reason is that the considered work is a hybrid \textit{discrete-continuous} action space problem and the altitude and scheduling are also closely coupled and should be carefully considered with the adjustments of RIS's phase-shift. Discretizing the altitude of the UAV and the phases of RIS elements into discrete actions could be one way to tackle this challenge. However, combining the three actions (UAV's altitude, scheduling policy and phases of RIS) into one single action space is still a challenge that needs to be addressed. The reason behind that, for a deployment with $M$ IoTDs, $\mathcal{K}$ discrete altitude actions and $Q$ discrete phases, a total of $(\mathcal{K} \times M \times F \times Q)$ possible actions need to be considered. In fact, efficient DRL algorithms are difficult or even often impossible to apply to solve large discrete action spaces since it increases the difficulty of learning. Since only one IoTD is scheduled per time slot, the phase-shift matrix for the RIS is properly  configured so that the reflected signals can be constructively added at  that IoTD. Due to that and to reduce the complexity of the learning and maintain a small size of the action space, we use the UAV altitude and scheduling policy as the main control objective. In other words, at each time-slot, the RL-agent (deployed at the BS) will adjust the altitude of the UAV and schedule a transmission along with an appropriate adjustment for the RIS's phase-shift according to the user of interest. Thus, the RL-agent executes two actions simultaneously. Specifically, the deployed UAV adjusts its altitude and at the same time, schedules an IoTD depending on the phases of RIS elements and the activation pattern. 
~\vspace{-0.2in}
\subsection{MDP Formulation} \label{State}

The considered problem  is first modeled as a Markov Decision Process (MDP). Afterward, a DRL based on PPO  algorithm is proposed for finding the control policy that governs both the UAV's altitude and the scheduling decision within unknown activation pattern. We define the  state $\cal{S}$, action $\cal{A}$, and reward $\cal{R}$ functions as follows:

1) \textbf{State} $\cal{S}$ in the PPO model:  The state at time-slot $n$ is defined as $s[n] =(\boldsymbol{A[n]}, \boldsymbol{\Upsilon[n]},  {H_U[n]})$, where $\boldsymbol{A[n]}=(A_1[n], ...., A_i[n], ....., A_M[n])$  and $\boldsymbol{\Upsilon[n]}=(\Upsilon_{1}[n],......\Upsilon_{i}[n],.....\Upsilon_{M}[n])$ is a vector of size $M$ containing the SNR at time-slot $n$ when the signals from different paths are coherently combined through the phases of RIS elements.

2) \textbf{Action} $\cal{A}$ in the PPO model:  At each step-slot $n$, the RL-agent  executes an action $a[n]$ denoted by $a[n] = (\boldsymbol{\xi[n]},  \boldsymbol{\kappa[n]})$, where $\boldsymbol{\xi[n]} \in \mathcal{W} =[\alpha_{1}[n] ,....\alpha_{i}[n] ,.....\alpha_{M}[n] $]. $\boldsymbol{\kappa[n]} \in \mathcal{K} = \{(1,0, 0),(0,1, 0),(0,0, 1)\}$, where $\kappa[n]$ represents the variable quantity of the altitude distance. $\boldsymbol{\kappa[n]} = (1,0,0)$  means the UAV adjusted its altitude upward at time-slot $n$; $\boldsymbol{\kappa[n]} = (0,1, 0)$ means the UAV adjusted its altitude downward at time-slot $n$; and $\boldsymbol{\kappa[n]} = (0,0, 1)$ means the UAV is hovering. Hence, the system action space at time-slot $n$ is $a[n] = (\boldsymbol{\xi[n]},  \boldsymbol{\kappa[n]}) \in \cal{A} = \mathcal{W} \times \mathcal{K}$. Thus, the RL-agent adjusts the UAV's altitude and decides which IoTD to transmit its status-update to the BS.

3) \textbf{Reward} $\cal{R}$ in the PPO algorithm: The immediate reward $r[n]$ is defined as a negative summation of AoI. Therefore, the RL-agent is motivated to minimize the AoI by optimizing scheduling decisions and altitude control of the UAV. 

Generally, MDP problems with predefined state transition probabilities can be solved using Dynamic Programming (DP). However, since the UAV is dispatched with no prior knowledge on the activation patterns of IoTDs, then DP  algorithm cannot be applied. Thus, to solve the considered problem a DRL based on PPO algorithm\footnote{It is noteworthy that, PPO demonstrates performance comparable to or better than state-of-the-art DRL approaches. Therefore, PPO has become the default reinforcement learning algorithm at OpenAI.} is employed in the next subsection.

\begin{algorithm}[!t]\small
		\caption{The PPO Based Framework.}
		\textbf{Initialize} $\theta$ randomly, $\pi_{\theta_{old}}$ with ${\theta_{old}}   \leftarrow \theta$ and $H_U[1]= H_S$.\\
		\For {Iteration = 0,1,··· }
		{ 
			\For {$l$= 0,1,.. $L$ }
			{    
				Get $(\boldsymbol{A[n]}, \boldsymbol{\Upsilon[n]},  {H_U[n]})$ from the environment.	\\
				Sample action $a[l] \sim \pi_{\theta_{old}}$.\\
				Take the action $a[l]$ that specifies the UAV's altitude and the IoTD scheduling.\\
                Configure $\boldsymbol{\Phi}[n]$ that maximizes the received SNR of the scheduled IoTD at the BS.\\
				\If {UAV violates the allowable altitude range}
				{
				Add a penalty, cancel the movement of UAV and update $s[l+1]$.
			}
				Get relevant reward $r[l]$ and  $s[l+1]$.\\
				Store ($s[l], a[l], r[l], s[l+1]$) as one transition in the experience replay.\\
				Compute advantage estimate.
			}
			\For{epoch= 0,1,.. }
			{ 
				Use Eq.(\ref{RewardgrrvntsSum}) to compute the PPO objective.\\
				Optimize the overall objective function and update the policy via stochastic gradient ascent with ADAM.\\
			}
			Synchronize the sampling policy with $\theta_{old} \leftarrow\theta$.\\
			Clear up the stored data.
		}
\end{algorithm}

\begin{figure*}[!t]
	\centering
	\hspace*{-0.9cm}
	\subfigure[Convergence.]{
		\includegraphics[height=1.3in]{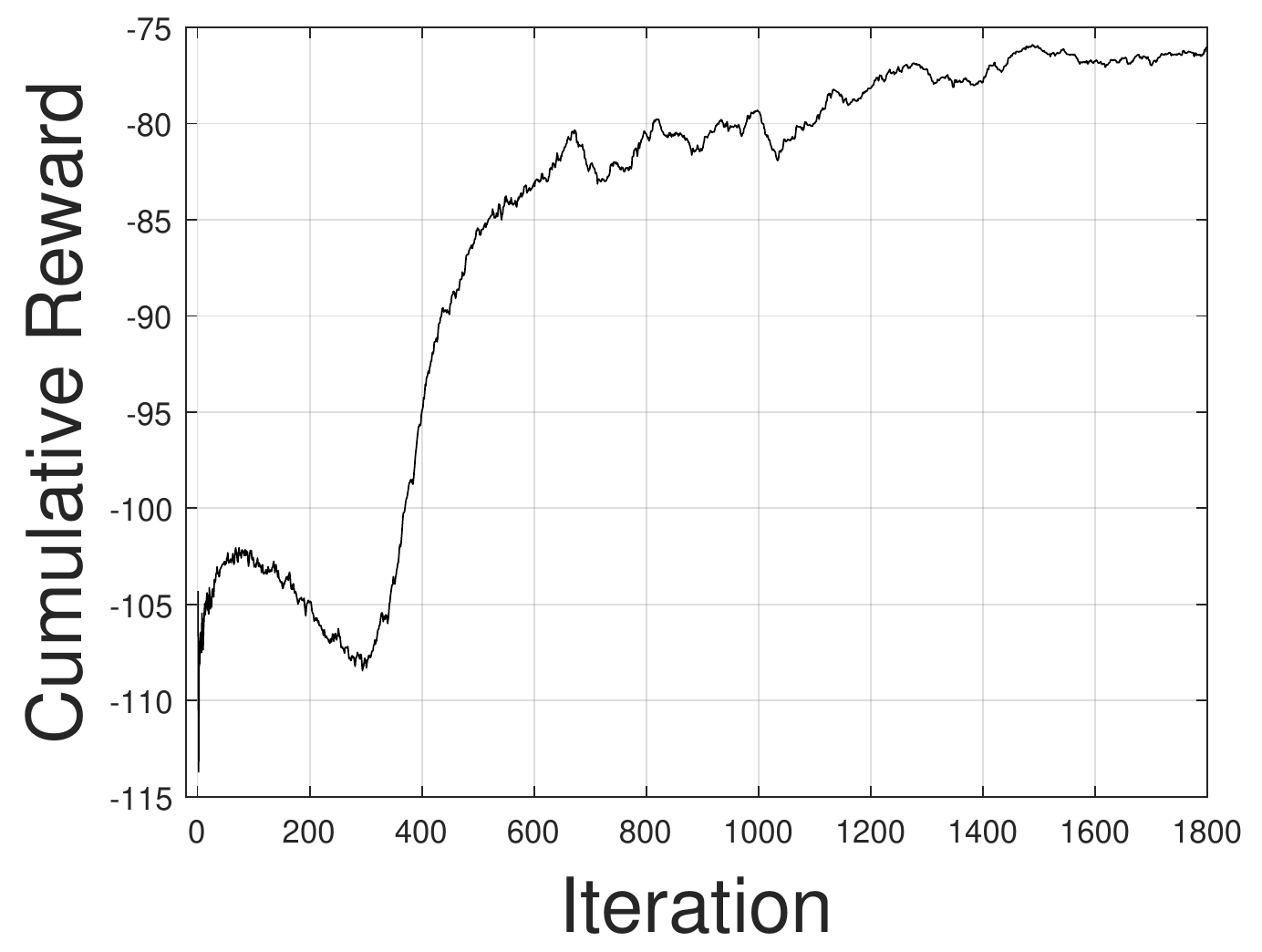}
		\label{subfig1}
			}
	\subfigure[Impact of number of IoTs.]{
		\includegraphics[height=1.3in]{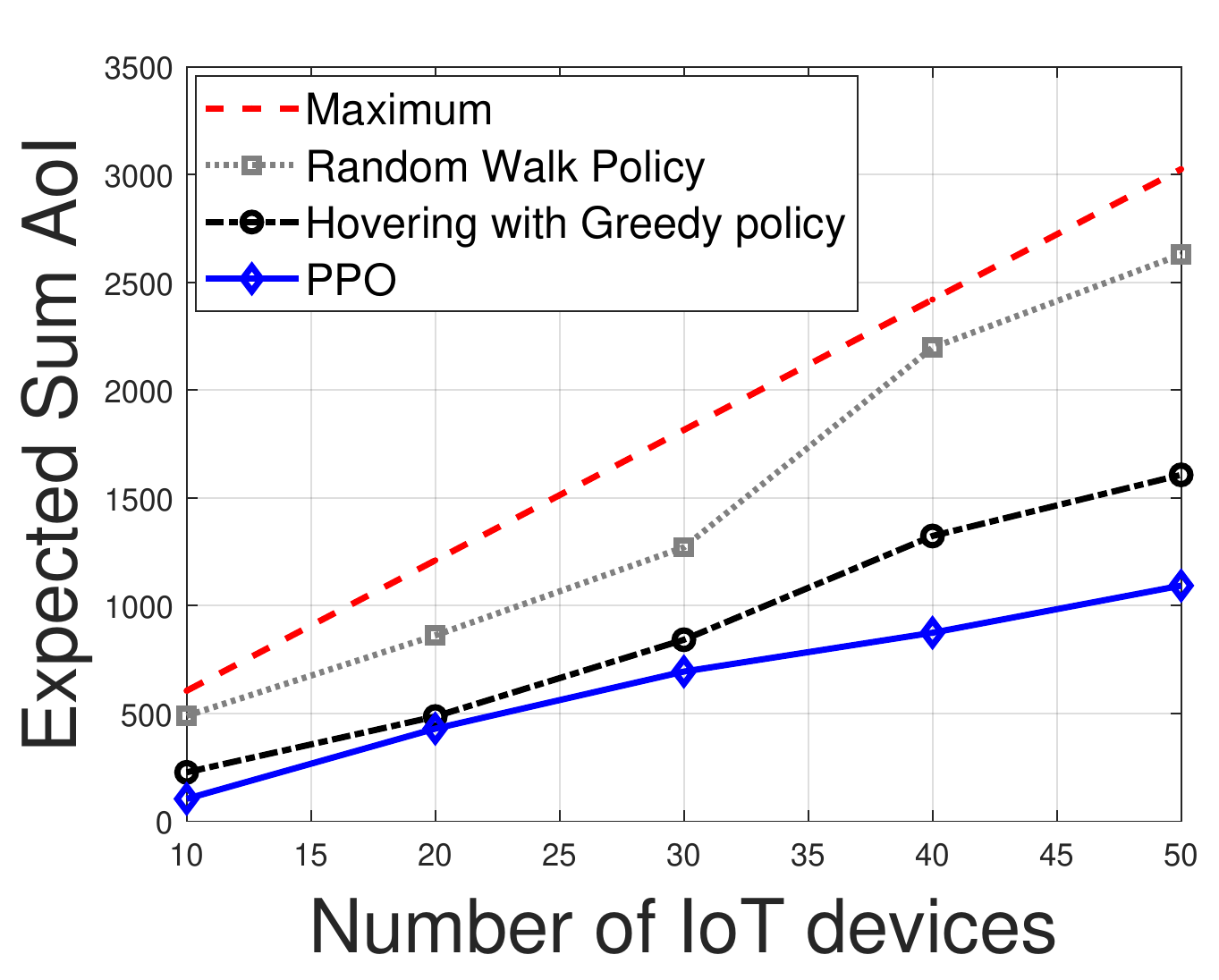}
		\label{fig:Resources1}
	}
	\subfigure[Average age per IoT device.]{
		\includegraphics[height=1.3in]{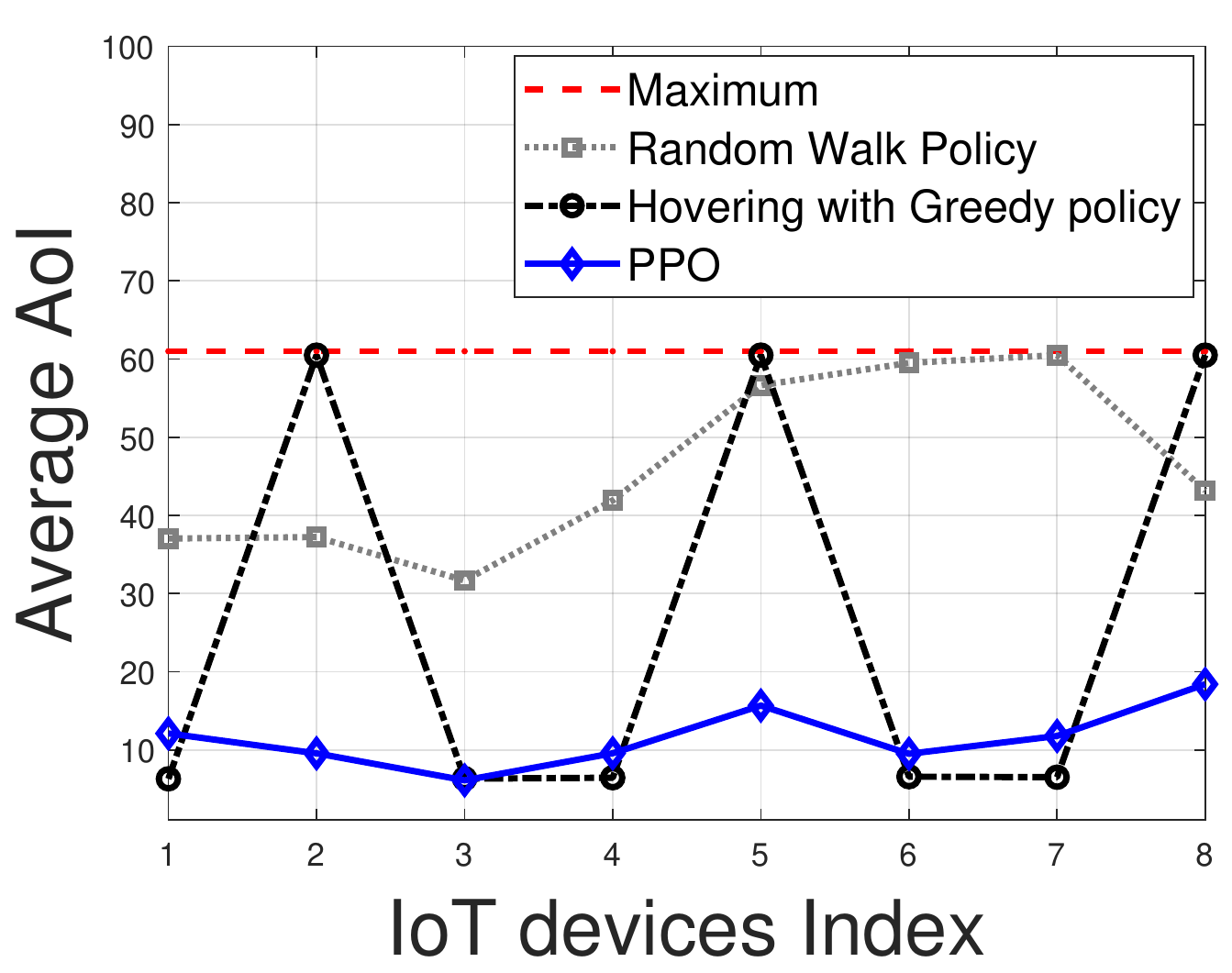}
		\label{fig:ReSpeed}
	}
	\subfigure[Impact of number of RIS elements.]{
		\includegraphics[height=1.3in]{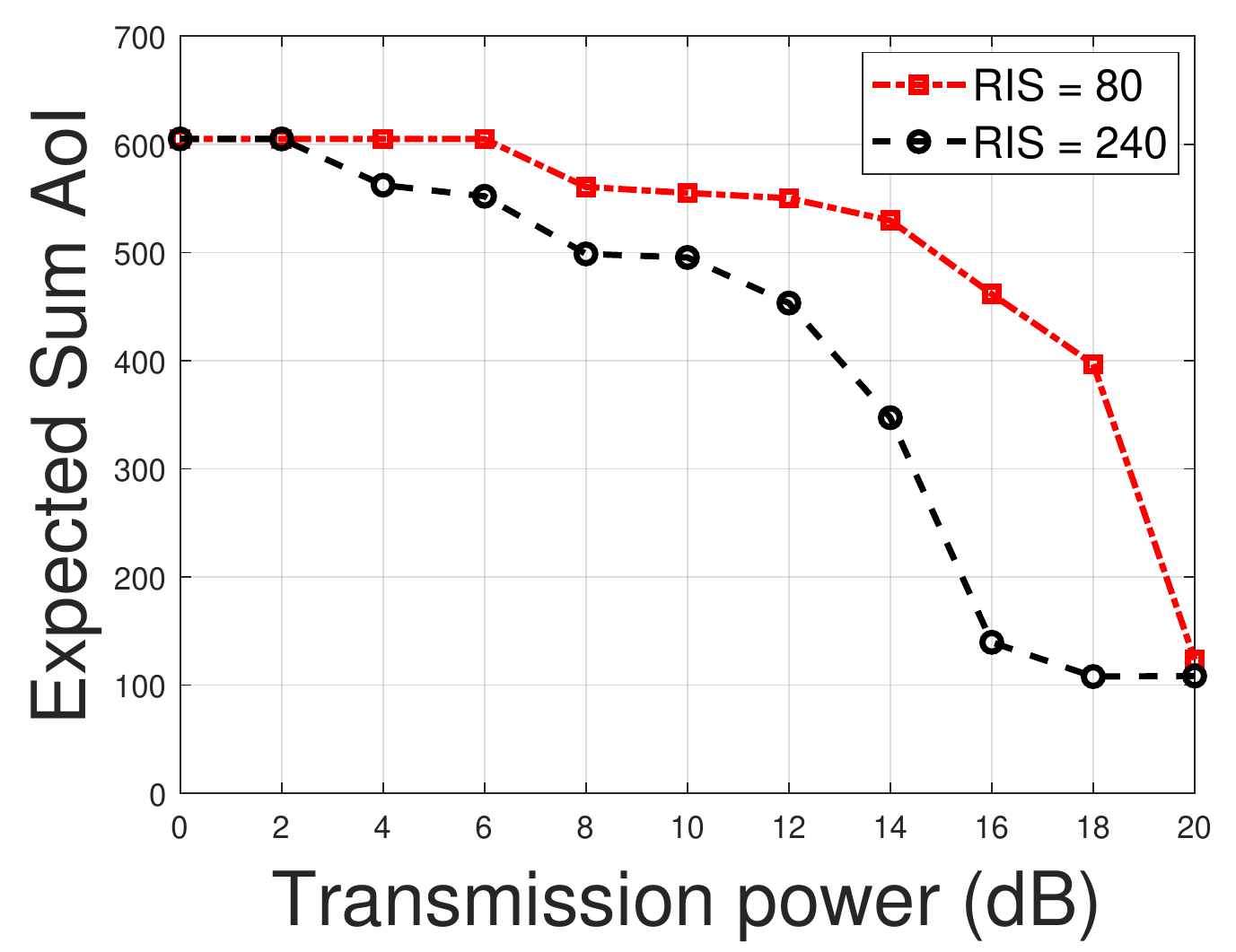}
		\label{fig:UAVs_Smin}
	}
	\caption{Optimizing the UAV altitude, scheduling policy and phases of RIS elements.}
	\label{test}
	\vspace{-1.5em}
\end{figure*}
~\vspace{-0.3in}
\subsection{Proposed Solution Description}\label{DRLnewrwww}

At the initialization stage (Line 1), the proposed algorithm randomly initializes the initial altitude of the UAV at $H_S$ to guarantee Eq.(\ref{eq33}). Besides, the deep neural networks (DNNs) parameter $\theta$ are randomly initialized, where DNNs have the same structure. In each training iteration, the PPO algorithm alternates between the sampling phase by running $L$ episodes (Lines 3-12) and the optimization/exploration phase (Lines 13-15). During each episode $l$, the current channel state information ${\textbf{h}}^{H}_{i \rightarrow U}[n]$ and ${\textbf{h}}_{U \rightarrow S}[n]$ are obtained. Then all possible phases of RIS elements $\boldsymbol{\Phi}[n]$ that achieve a coherent combination with the signals from different paths at the BS are obtained. The RL-agent (Line 4) then get observations $\boldsymbol{A[n]}$, $\boldsymbol{\Upsilon[n]}$ and ${H_U[n]}$ from environment at each time-slot. The UAV then executes action $a[l]$ from the policy $\pi_{\theta_{old}}$ (Lines 6-7), where the sampled action, $a[l]$, represents the current altitude of the UAV and the scheduling policy according to the phases of RIS configuration for each IoTD that achieve the maximum SNR. Eq.(\ref{eC22443}), Eq.(\ref{AOIgvgatsdsdion}) and Eq.(\ref{AOIgvgatisdson}) are implicitly defined in Lines 4-7, where the states of the MDP and actions are defined. In this step, the PPO algorithm assigns the binary value "1" to the scheduled IoTD to transmit its status-update and assign "0" for other IoTDs to guarantee Eq.(\ref{eC1443}) and Eq.(\ref{eq:226263dcd22})	.

During each episode (Lines 8-9), the RL-agent guides the UAV to avoid the action that violates the altitude constraint (i.e., Eq.(\ref{eq11})) by abandoning the corresponding altitude action and apply a penalty to the reward. After taking the current action, the UAV receives the relevant reward (Line 10), which represents the sum of the AoI for all IoTDs. After, the tuple of RL trajectory data of episode $l$, ($s[l], a[l], r[l], s[l+1])$ are buffered for the next phase (Line 11), then advantage estimate is computed (Line 12) to achieve efficient training of the policy. To update policy network $\pi_{\theta}$, the PPO clip objective function in each epoch is computed according to the below 
\allowdisplaybreaks
\begingroup
\begin{equation}\small
\begin{split}
\label{RewardgrrvntsSum}
&L^{CLIP}({\theta}) = {\E}_n \Bigg[\min\Big(\dfrac{{\pi}_{\theta} (a[n]|s[n])}{{\pi}_{\theta_{old}} (a[n]|s[n])} \mathfrak{A}_{\pi_{\theta_{old}}} (s[n],a[n]), \cr & {clip}\Big(\dfrac{{\pi}_{\theta} (a[n]|s[n])}{{\pi}_{\theta_{old}} (a[n]|s[n])} , 1+\epsilon, 1- \epsilon\Big) \mathfrak{A}_{\pi_{\theta_{old}}}(s[n],a[n])\Big)\Bigg],
\end{split}
\end{equation}
\endgroup
where $\epsilon$ is the clip fraction used to control the clip range. $\mathfrak{A}(s[n], a[n])$ is the advantage estimate in time-slot $n$ that is used to mitigate the high variance of the gradient. Then the PPO overall objective function is then optimized via stochastic gradient ascent (SGA) with ADAM. Finally, the policy $\pi_{\theta_{old}}$ is updated with the policy $\pi_{\theta}$ and the buffered data are dropped (Line 17) then a new iteration begins.

The complexity of the proposed algorithm can be expressed as the number of multiplications: $O(\sum_{p=1}^{P-1} n_p.n_{p-1})$, where $n_p$ is the number of neural units in the $p$-th hidden layer. The convergence of the algorithm is limited to simulations since it is hard to be analytically analyzed.

\section {Simulation And Numerical Analysis} \label{simulationgNEW}
We consider a square area of $0.5$ km $\times$ $0.5$ km, where IoTDs are distributed randomly. An UAV equipped with RIS is placed at the center of the area to relay the status-update information from IoTDs to the BS located at (2000, 500, 25) m. We assume that all the IoTDs have the same power budget. The frame duration $N$ is take as $N = 120$ while the communication parameters are taken as: $P = 20$ dBm, the path-loss exponent $\eta = 2.3$, the channel gain $\gamma_{0} = -20$ dBm, $\sigma^{2} = -110$ dBm, $\Upsilon_{th} = 0$ dB \cite{VPOORAGE} and $K_1 = K_2 = 8$ dB. The UAV altitude parameters are assumed as $H_S=100$ m, $H_{min}=10$ m, $H_{max}=1000$ m and $D_{max} =10$ m/s. The activation pattern of each IoTD is randomly generated according to Uniform distribution\footnote{The same solution approach  can be applied to any activation distribution.}. Table \ref{my-label} provides PPO hyperparameters. The results are collected by utilizing PyTorch library after 240K samples, where each sample corresponds to a snapshot of the network at a particular time-slot. 
%
%
%
\par We first verify the convergence of the proposed PPO algorithm in  Fig. \ref{subfig1}. It can be found that trained RL-agent can significantly improve the defined reward. This improvement begins to diminish when the RL-agent is well trained about the activation patterns of IoTDs and it starts to effectively adapt the UAVs’ altitude and the scheduling policy that minimizes the ESA. It is observed that our PPO algorithm converges to the steady point quite quickly under a reasonable choice of the hyperparameters. This indicates that our proposed algorithm deals effectively with incomplete knowledge of the activation pattern of the IoTDs.
\par To evaluate the effectiveness of the proposed algorithm, we develop two baselines policies as follows: 1) a random walk policy which randomly selects an IoTD to relay its status-update information along with adjusting the phases of RIS so that the reflected signals can be constructively added at the selected  IoTD while changing the altitude of UAV randomly, and 2) hovering with the greedy policy where the UAV iteratively searches for the best height that satisfies the reliability constraint for the most IoTDs. Then, the UAV selects the IoTD with the maximum current AoI. Similar to the random walk policy, the phase-shift matrix of the RIS is adjusted with the same way. The baseline policies are adequate policies since the former policy explores all possible actions, thus, may obtain some actions that result in decreasing the AoI, while the latter policy is heuristically a good policy since it always selects the IoTDs with higher AoI to relay their status.
\par Fig. \ref{fig:Resources1} illustrates the impact of the number of IoTDs on the PPO algorithm compared to the baselines policies. It can be noted that the proposed algorithm is able to minimize the ESA for a lower number of IoTDs since each IoTD enjoys more frequent scheduling. However, for a large number of IoTDs we can see that the ESA increases since more scheduling is needed to decrease the ESA. Besides, the hovering with the greedy policy is more effective than the random walk policy since it always selects the IoTD with the highest AoI value. We also can observe that the proposed algorithm outperforms all the baselines. This
is expected since the baselines are unable to learn the activation pattern of the IoTDs and the altitude adaptation of the UAV is not considered in the baselines.
\begin{table}[t]\small
	\centering
	\caption{Simulation Parameters}
	\label{my-label}
	\small\addtolength{\tabcolsep}{-5 pt}
	\scalebox{0.9}{
	\begin{tabular}{|c|c|c|c|}
		\hline
		\centering	\textbf{Parameter}   & \textbf{Value}  \\ \hline
		\hline
		\centering	Activation Functions   &    Softmax and Tanh      \\ \hline
		\centering	Learning Rate   &   $0.001$       \\ \hline
		\centering	Reward Discount   &    $0.9$      \\ \hline
		\centering	Number of Hidden Layers for Networks  &    $3$  \\ \hline
		\centering	Number of Neurons   &    $64$      \\ \hline
		\centering  Loss Coefficients $K_1$ and $K_2$  &    $0.5$ and $0.01$   \\ \hline
		\centering  Update Policy Length, L&  $240$      \\ \hline
		\centering  Clip Fraction, $\epsilon$&  $0.2$      \\ \hline
		\centering	Optimizer Technique   &     Adam      \\ \hline
	\end{tabular}}
\end{table}
\par We plot the average age as another performance metric for a set of IoTDs in Fig. \ref{fig:ReSpeed}. The average age of IoTD $i$ within time $N$ is calculated by $\frac{1}{N}\sum_{n=0}^N  A_i^n, \forall i$. Obviously, it can be seen that the proposed PPO has a lower average sum AoI per IoTD compared to the baseline policies. The average age gap among the policies is relatively high, which demonstrates the importance of learning of the activation pattern of IoTD and adjust the altitude of the UAV with communication scheduling. Furthermore, the  hovering  with  the greedy  policy, on the one hand, significantly decreases the AoI for some IoTDs. On the other hand, it increases the AoI to the maximum for other IoTDs. This is because the the  hovering  with  the greedy  policy only schedules transmission for IoTDs that satisfy the SNR threshold. This finding justifies the robustness of the proposed PPO algorithm in terms of minimizing average AoI. 

\par Finally, in  Fig. \ref{fig:UAVs_Smin}, we show that both the power budget of the IoTDs and the number of RIS's elements have a great impact on the ESA. Increasing the transmit power of IoTD leads to a direct enhancement of the achieved SNR at the BS. However, this may not be allowable in certain IoT applications. As a result, it is better to increase the number of reflecting elements per RIS which results in enhancing the quality of the communication link between the IoTD and the BS, which in turns improves the achieved SNR and the ESA.
~\vspace{-0.1in}
\section {Conclusion} \label{concluNewsions}
This letter proposed a new relaying system to maintain the freshness of information of remote Internet of Things wireless network by integrating the unmanned aerial vehicle (UAV) and the reconfigurable intelligent surfaces (RIS). The altitude of the UAV, the transmission scheduling, and phase-shift matrix of RIS elements are optimized to minimize the expected sum \textit{Age-of-Information}. To tackle this mixed-integer non-convex problem, proximal policy optimization algorithm is proposed. Numerical results demonstrate that the proposed algorithm can significantly minimize the AoI compared to other baselines such as random walk and heuristic greedy algorithms.

\bibliographystyle{IEEEtran}
\bibliography{IEEEabrv,reference}

\end{document}